# Analysis of the time series in the space maser signals

**S. Siparov**
Department of Physics, State University of Civil Aviation, St-Petersburg, RF
National Research University of Information Technologies, Mechanics and Optics, St-Petersburg, RF
**V. Samodurov**
Radio Astronomy Observatory of RAS, Puschino, RF
**G. Laptev**
Teknavo Group, St-Petersburg, RF

We analyze the data of the observations of the radio sources frequently found in space. They are believed to be the sets of molecular condensations each of which works as a maser, so that the whole set produces a characteristic spectrum. It turns out that in some cases the intensity of one of the components of such spectrum corresponding to a single condensation changes periodically with a period of dozens of minutes or of hours.

Space masers are the sources of almost monochromatic radiation produced by the clouds of molecules in meta-stable states. This radiation can be registered by radio telescopes. A single session of observation usually lasts for several hours and every small period of time the gathered spectrum is transferred to the recorder. The effect we discuss here is a periodic change with time of a certain single detail on the spectrum while the other details do not change in the similar way (at least visually) during the session. Such selective behavior means that this is not an instrumental, weather or interstellar medium effect. Therefore, it seemed reasonable to investigate the corresponding time dependencies for various space masers.

The specially modernized registration channel of the telescope provided the gathering of the data for the accumulation period equal to 30 seconds. Every half an hour during the session the observations were interrupted for 5 minutes in order to calibrate the antenna. Preliminary signal processing included the instrumental control (antenna and registering channel) and all the due averaging intended to eliminate the weather and instability effects (see [1]). The special program recently designed for the processing of the obtained data does the following. At first, it sums up the flux values for corresponding frequencies for all the spectra obtained during the given session and finds out the frequencies corresponding to all the local minima in this sum. Then it takes every region between the neighboring minima (e.g. see the detail between the dashed lines on FIG.1) for every flux spectrum, follows its change with time during the given observation session, and produces the corresponding time series for this region, FIG.2. The analysis of the time series includes three stages: fast Fourier transformation (FFT) which is used in order to discover the presence of a periodic component corresponding to a stand-alone peak, FIG.3; Lomb-Scargle procedure (LS) which exploits a more complicated algorithm in order to construct a periodogram – a plot with a set of peaks corresponding to the found periodic components, FIG.4; and, finally, the modified LS algorithm [2] which filters out the periodic features characterizing the observation procedure itself, FIG.5. The FFT was already used in [1] in order to show that the periodic components can be found in space masers signal, but this approach is limited because it deals only with equidistant points and requires some interpolation.

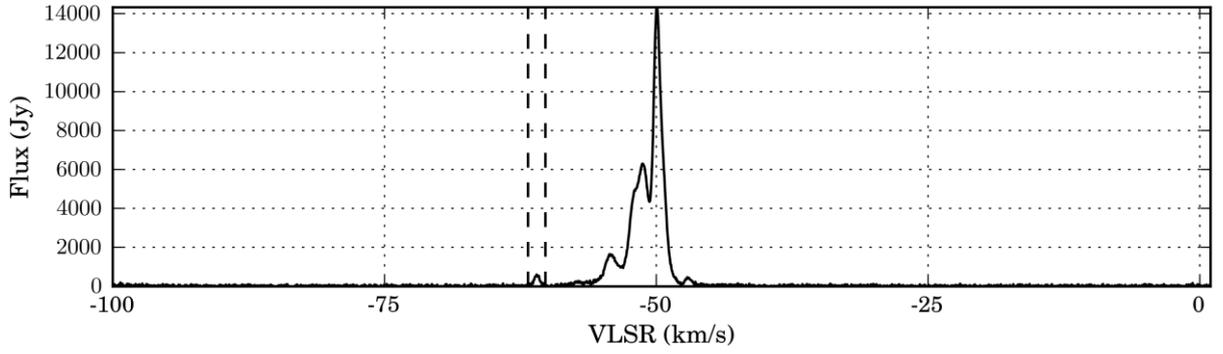

FIG.1. Maser W3(OH): RA 2h23m18s, Dec +61$^0$38'58''. (RT-22, June 30, 2009, Puschino RAO RAS)

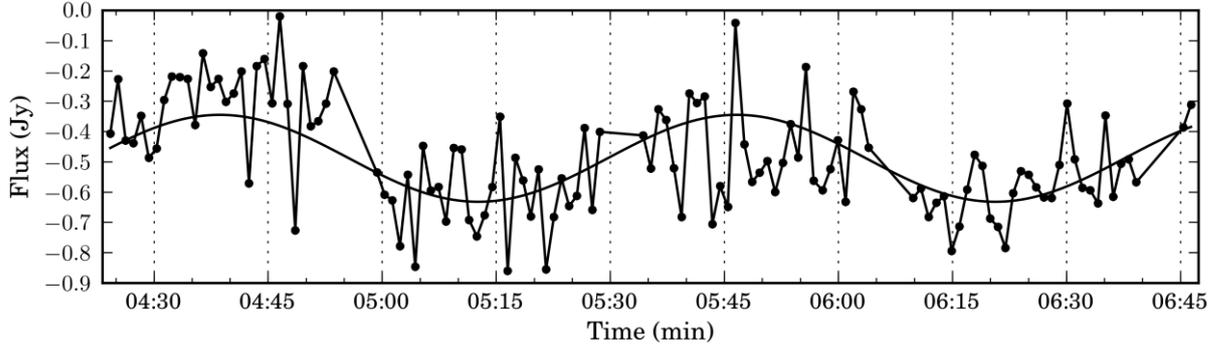

FIG.2. Time dependence of the selected detail shown on FIG.1

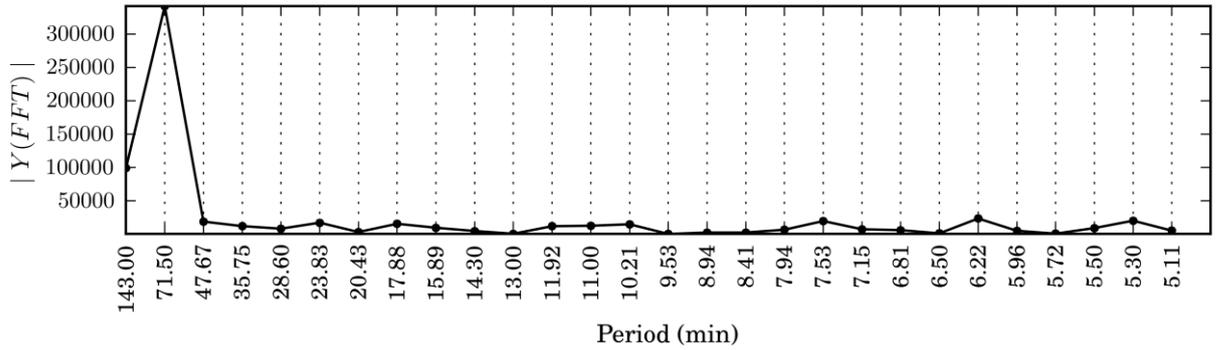

FIG.3. Preliminary FFT analysis of the time dependence

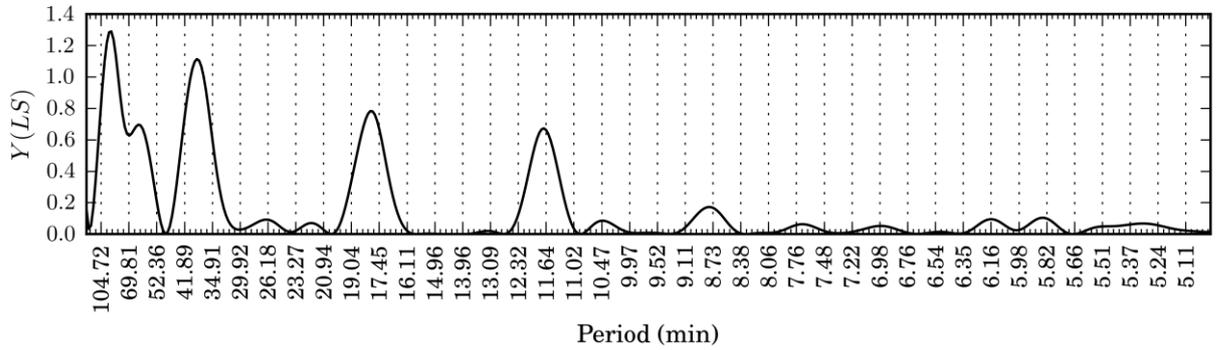

FIG.4. Lomb-Scargle periodogram

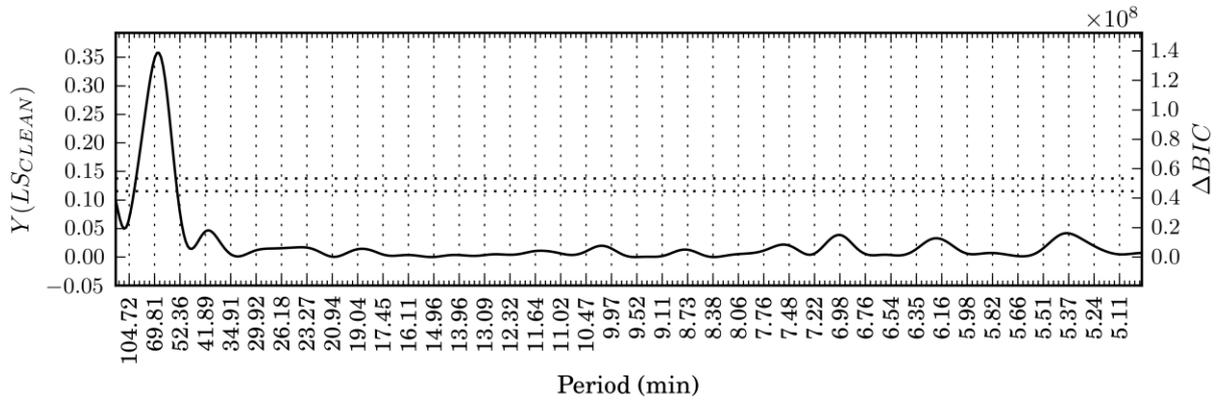

FIG.5. Lomb-Scargle periodogram with window function filter.

On the periodogram (FIG.4) one can see *several* peaks, but some of them could be due to the total time of the session, to the period of antenna calibration and to the number of measurements between calibrations [3]. High peak on FIG.5 presents a period of 68 minutes corresponding to the non-stationary component of the selected maser signal detail marked on FIG.1. It should be underlined that all the other details of the spectrum present on FIG.1 do not periodically change and do not lead to a peak like that on FIG.5. Therefore, this peak characterizes the periodical change of the intensity of radiation emitted by only one of the condensations that form the space maser cloud.

The same procedure was performed for about 150 sessions of observations of 49 radio sources presenting $H_2O$ space masers. Signals like that given on FIG.5 were also found for the following sources: Cep A, IRAS 16293-2422, Ori A, RT Vir, W 49 N, W 31 A, W 3 (2), VY CMa. The detailed report is coming soon.

Space masers are known to reveal not only long flux variations but also short and ultra-short ones [4-5]. But ultra-short *periodic* flux variations of *separate components* were reported only in [1]. The local physical mechanism belonging to the vicinity of a cloud that could provide such effect is unclear. Interpretation given in [1] deals with the so-called optic-metrical parametric resonance (OMPR), and now it seems still valid. The theory of OMPR is given in [6-7] and in more detail in [8]. The essence of it is as follows. When a quantum object like an atom or a molecule that can be described by the two-level model (TLA) is put into the spectroscopically strong quasi-resonant electromagnetic field, there appears a new parameter, which is called Rabi frequency, and it characterizes the strength of the field. It corresponds to the frequency of the population oscillation between the upper and the lower levels of the TLA due to the stimulated transitions. If the distance between the radiating atom and the radiation receiver varies with the frequency related to Rabi frequency and some other conditions are fulfilled, the parametric resonance is possible, and there appears a non-stationary component in the registered radiation spectrum. 'Non-stationary' here means that the received spectrum contains a component with a specified frequency whose intensity periodically changes. Space maser is an object whose properties suffice the conditions needed for the OMPR effect: space maser particles can be described by the two-level model; the saturated maser provides the strong field; and if there is a *periodical* gravitational wave (GW) falling upon the maser at due direction, the distance between maser's particles and the observer on the Earth periodically changes. The specific conditions needed for the OMPR that are given in [6] can be fulfilled for the concrete cases [7-8]. It was shown that the action of the GW on the atomic levels is negligibly small in comparison with its action upon the radiation itself and upon the distance to the receiver. The solution of the Bloch's equations for the OMPR problem has the form

$$\text{Im}(\rho_{21}) \sim \frac{\alpha_1}{D} \cos 2Dt + O(\varepsilon)$$

Here $\rho_{21}$ is the density matrix component, $\alpha_1 = \mu E/\hbar$ is Rabi frequency ($\mu$ is induced dipole moment, $\hbar$ is Planck constant, $E$ is the electric stress of the field), $D$ is the frequency of GW acting on a space maser, $\varepsilon = \gamma/\alpha_1$ characterizes the maser field and it is a small value for the strong field, $\gamma$ is the decay rate of the excited state (for $H_2O$-space maser $\gamma \sim 10^{-10}$Hz). Notice, that contrary to the first order effects that are usually discussed when speaking about the GW detection, the amplitude of the non-stationary component does not depend on the amplitude of the GW, and it is of the same order as the signal itself. This is due to the parametric resonance.

Thus, if we use this approach in order to find the traces of the GW, the main problem is to find the proper astrophysical system, i.e. the configuration of a maser, a short-period (tens of minutes) binary star that could be the source of the GW, and an observer on the Earth. Simple considerations show that if the Earth and the GW source are on the ends of a sphere diameter, the maser should be located close to the surface of this sphere. And if the Earth and the maser are the top of a cone and the center of its base, the GW source should be located close to the plane of the cone's base. Therefore, if the location of the maser is known and the analysis of the time series of its signal gives the frequency of the non-stationary component (which is equal to $2D$), we could try to look for a close binary with the orbital period corresponding to $D$, which suffices the geometrical configuration mentioned above.

Here we have the W3(OH) maser source with coordinates on the sky given on FIG.1, the distance to it is about 2000 pc [9], and the found period is 68 minutes. Thus, the binary with the demanded period of 136 minutes could be Cyg v2214 (or KPD 1930+2752) [10], whose coordinates are RA 19h32m14.81s, Dec +27$^0$58'35.5'', and whose distance to the Earth is about 500 pc [11]. The geometrical configuration is not perfect, but the demanded geometrical conditions are sufficed well enough. On FIG.6 there is the peak that was obtained by the similar analysis for the observation session 4 days before the session investigated on the previous figures. In order to check the phase coincidence between the signals registered on June 26 and June 30, the corresponding time dependencies were approximated by the sine functions. Then the time difference (in minutes) between a maximum approximating the plot on FIG.2 and a maximum on the similar plot for June 26 session was divided by 68. The result was integer within the accuracy of 0.029, which seems good enough to support the OMPR interpretation.

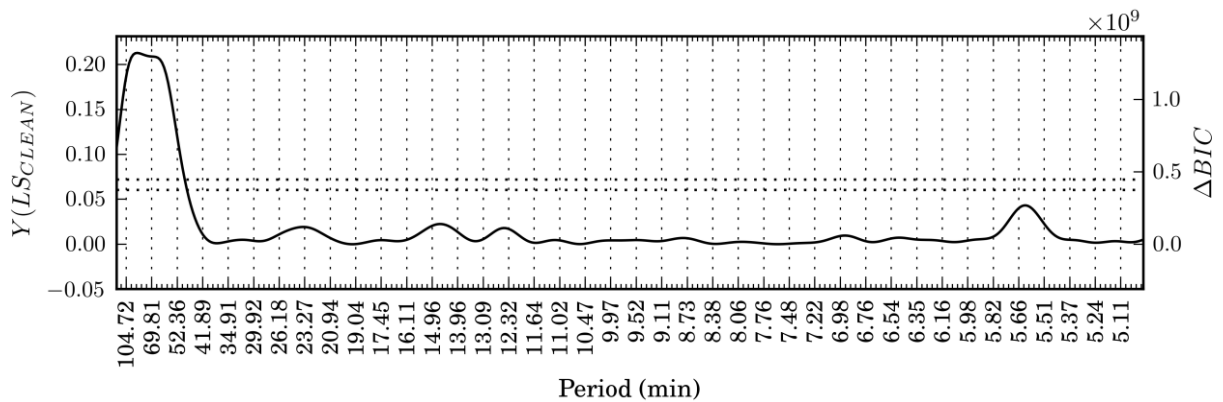

FIG.6. Lomb-Scargle periodogram with window function filter for maser W3(OH): RA 2h23m18s, Dec +61$^0$38'58''. (RT-22, June 26, 2009, Puschino RAO RAS)

It should be mentioned that since a space maser is a dynamical object, the OMPR conditions for a condensation could be fulfilled not permanently but from time to time.

**Acknowledgement**

# References

[1] – S.V.Siparov, V.A.Samodurov. **Comp. Optics** 33 (1), p.79, 2009 (rus); (arXiv:0904.1875 [astro-ph], 2009 (eng))

[2] – http://www.astroml.org.

[3] – V.Vitiazev. *Analisis of irregular time series*, SPb State University, St-Petersburg, 2001

[4] – A.M.S.Richards et al. **Astrophys. and Space Science** 295, p. 19, 2005

[5] – V.A.Samodurov et al. **Astron. and Astrophys. Transactions** 25, No.5, p. 393–398, 2006

[6] – S.Siparov. **Astronomy and Astrophysics** 416, p. 815-824, 2004

[7] – S.Siparov. **Astron. and Astrophys. Transactions** 27, p. 1–7, 2010

[8] – S.Siparov. *Introduction to the Anisotropic Geometrodynamics*, World Scientific, New Jersey – London – Singapore, 2011.

[9] – http://www3.mpifr-bonn.mpg.de/staff/abrunthaler/pub/w3oh.pdf

[10] – The Binary Star Database, http://bdb.inasan.ru/

[11] – http://news.sciencemag.org/2000/07/waiting-big-bang